\documentclass[useAMS]{mn2e}
\usepackage{times,epsfig} 
\makeatletter
\def\figsize{\ifSFB@referee0.5\hsize\else\hsize\fi}
\makeatother

\def\DS     {\displaystyle}
\def\mic    {\hbox{$\mu$m}}
\def\eq#1{\begin{equation} #1 \end{equation}}
\def\E#1{\hbox{$10^{#1}$}}
\def\sub#1{_{\rm #1}}
\def\about  {\hbox{$\sim$}}
\def\x      {\hbox{$\times$}}
\def\half   {\hbox{$1\over2$}}
\def\Mo     {\hbox{M$_\odot$}}
\def\Lo     {\hbox{L$_\odot$}}
\def\tV     {\hbox{$\tau\sub V$}}
\def\deg    {\hbox{$^\circ$}}
\def\T      {\hbox{$T_\star$}}
\def\ts     {\hbox{$\theta_{\rm s}$}}
\def\ta     {\hbox{$\theta_{\rm a}$}}
\def\R      {\hbox{$R_\star$}}
\def\Rd     {\hbox{$R_{\rm d}$}}
\def\Rin    {\hbox{$R_{\rm in}$}}
\def\Yd     {\hbox{$Y_{\rm d}$}}
\def\Td     {\hbox{$T_{\rm d}$}}
\def\ra     {\hbox{$r_{\rm a}$}}
\def\Rh     {\hbox{$R_{\rm h}$}}
\def\Rs     {\hbox{$R_{\rm s}$}}
\def\Rss    {\hbox{$R_{\rm s}^2$}}
\def\Fs     {\hbox{$F_{\rm s}$}}
\def\Ts     {\hbox{$T_{\rm s}$}}
\def\Ldisk  {\hbox{$L_{\rm disk}$}}
\def\Lsph   {\hbox{$L_{\rm sph}$}}
\def\Fdisk  {\hbox{$F_{\rm disk}$}}
\def\Fsph   {\hbox{$F_{\rm sph}$}}
\def\Tout   {\hbox{$T\sub{out}$}}
\def\lout   {\hbox{$\lambda\sub{out}$}}
\def\H      {\hbox{$\cal H$}}
\def\N      {\hbox{$\cal N$}}
\def\n      {\hbox{$n_{\rm d}$}}

\newcount\hour
\newcount\minute
\def\clock    {\hour = \time \divide\hour by 60 \minute = \hour
                 \multiply\minute by -60  \advance\minute by \time
                 \number\hour:\ifnum\minute < 10 {0\number\minute}
                                              \else\number\minute
                                              \fi}
\def\Draft{Submitted September 16, 2002; revised August 19, 2003;
 accepted August 25, 2003}

\title[Disks and Halos in PMS Stars]
{Disks and Halos in Pre-Main-Sequence Stars}

\author[Vinkovi\'c et al.]
   {Dejan Vinkovi\'{c},$^1$ \v{Z}eljko Ivezi\'{c},$^2$
    Anatoly S. Miroshnichenko$^3$ and Moshe Elitzur$^1$\\
$^1$Department of Physics \& Astronomy, University of Kentucky,
    Lexington, KY 40506, USA; dejan@pa.uky.edu, moshe@uky.edu\\
$^2$Department of Astrophysical Sciences, Princeton University,
    Princeton, NJ 08544, USA; ivezic@astro.Princeton.edu;
    H.N.\ Russel Fellow \\
$^3$Department of Physics \& Astronomy,
    University of Toledo, Toledo, OH 43606, USA;
    anatoly@physics.utoledo.edu
}

\date{\Draft}
\pagerange{\pageref{firstpage}--\pageref{lastpage}} \pubyear{2003}

\begin{document}
\label{firstpage}

\maketitle

\begin{abstract}
We study the IR emission from flared disks with and without additional
optically thin halos. Flux calculations of a flared disk in vacuum can be
considered a special case of the more general family of models in which the
disk is imbedded in an optically thin halo. In the absence of such halo, flux
measurements can never rule out its existence because the disk flaring surface
defines a mathematically equivalent halo that produces the exact same flux at
all IR wavelengths. When a flared disk with height $H$ at its outer radius $R$
is imbedded in a halo whose optical depth at visual wavelengths is
$\tau\sub{halo}$, the system IR flux is dominated by the halo whenever
$\tau\sub{halo} > \frac14H/R$. Even when its optical depth is much smaller, the
halo can still have a significant effect on the disk temperature profile.
Imaging is the only way to rule out the existence of a potential halo, and we
identify a decisive test that extracts a signature unique to flared disks from
imaging observations.
\end{abstract}

\begin{keywords}
circumstellar matter --- dust --- infrared: stars --- radiative transfer ---
stars: imaging --- stars: formation --- stars: pre-main-sequence
\end{keywords}

\section{INTRODUCTION}

Modeling the IR radiation of pre-main-sequence stars has traditionally involved
``classic'' geometrically-thin optically-thick disks. This approach fails to
produce many features of observed spectral energy distributions (SED) in both T
Tauri stars (TTS) and Herbig Ae/Be stars (Haebes). One way out of this problem
is to supplement the disk emission with a surrounding optically thin halo
(e.g.\ Butner, Natta \& Evans 1994; Miroshnichenko et al 1999, MIVE hereafter).
A simpler alternative was proposed by Chiang \& Goldreich (1997, CG hereafter):
The surface skin of any optically thick object is of course optically thin. The
emission from the disk surface layer can become significant under certain
flaring conditions, and CG present SED fits of a number of TTS purely in terms
of flared disks. This proposal was extended to SED modeling of Haebes by Chiang
at al (2001) and Natta et al (2001).

While the CG model successfully solves the SED problem with a simple physical
explanation, recent high-resolution imaging observations of both TTS and Haebes
reveal the presence of halos, ignored in the CG approach. Most striking is the
case of GM Aur, a classical TTS whose SED modeling was presented as evidence of
the success of the CG approach (Chiang \& Goldreich 1999). Indeed, HST/NICMOS
images by Schneider et al (2003) show a flared disk with radius of \about300
AU, but they also reveal the presence of a surrounding tenuous envelope
extending to a radius of \about1000 AU. Using detailed Monte Carlo
calculations, Schneider et al find that simplified models with flared disk
without an imbedding halo fail to replicate the scattered light intensity
pattern seen in the NICMOS images. The addition of a halo is essential for
successful modeling of both the SED and the images. Similarly, HST and
ground-based imaging by Stapelfeldt et al (2003) of the TTS HV Tau C show a
nearly edge-on flared disk imbedded in a more extended nebulosity. From
detailed model calculations Stapelfeldt et al find that although it is possible
to fit the SED purely with a flared disk, the flaring is unreasonably large and
such models do not reproduce the image adequately. They conclude that a flared
disk alone is an inadequate model and that an additional, extended component to
the circumstellar density distribution is needed to explain the observations.
Finally, from detailed modeling of flared disk SED Kikuchi, Nakamoto \& Ogochi
(2002) conclude that halos are necessary supplements in explaining
flat-spectrum TTS.

High resolution observations of Haebes give similar evidence for imbedding
halos. Combining space- and ground-based observations of HD 100546, the nearest
Herbig Be star, Grady et al (2001) resolve the disk, which extends to 5\arcsec\
(\about 500 AU) from the star and displays the elongation of inclined viewing
with $i$ = 49\deg. But they also find that the disk is imbedded in a more
extended halo (\about 10\arcsec\ radius) that is roughly circular in shape and
optically thin (background stars are visible through it). Polomski et al (2002)
performed high-resolution ground-based observations of a sample of Haebes whose
mid-IR emission was claimed to be disk-dominated by Hillenbrand et al (1992).
They find instead that the emission is not confined to an optically thick disk
but originates in a more complex environment that includes large, extended dust
envelopes. From analysis of ISO data, Creech-Eakman et al (2002) conclude that
only a subset of the Haebes they observe can be described purely in terms of
the CG model (notably, E.I.\ Chiang is a coauthor of this study).

How can the SED of the same T-Tauri star, GM Aur, be fitted successfully with a
flared disk both with a halo (Schneider et al 2003) and without one (Chiang \&
Goldreich 1999)? Is it at all possible to distinguish between these two cases?
What, if any, is the unique radiative signature of each component? Obviously,
imbedding a disk in a tenuous halo with a very small optical depth is not going
to affect its emission appreciably. At which stage then does the halo assume a
significant role? Here we address these questions. In appendix A we derive
general results for the emission from optically thin dust in an arbitrary
geometry. We apply these results to the surface layers of flared disks in \S2
and to halos in which the disks could be imbedded in \S3.

\section{FLARED DISKS}

\subsection{The CG layer and its equivalence to spherical halo}
\label{sec:CG}

\begin{figure}
\centering \leavevmode
 \psfig{file=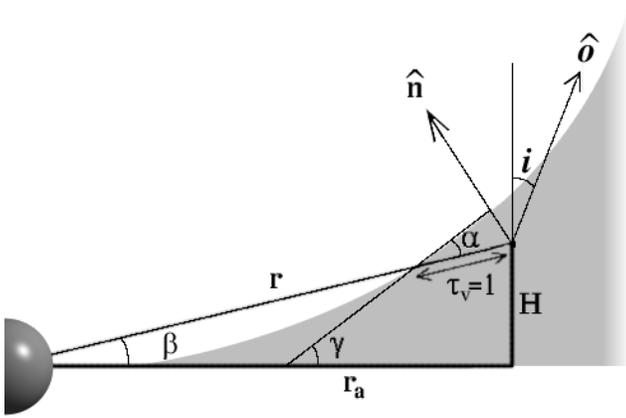,width=\figsize,clip}
\caption{Model geometry and notations for a flared disk and its CG surface
layer. The unit vector $\hat{o}$ points at the observer's direction.}
\label{fig:CG}
\end{figure}

Chiang and Goldreich (1997) noted that the emission from the optically thin
surface layer of an optically thick disk, which has been neglected
traditionally, can become significant under certain flaring conditions. The
stellar radiation penetrates to an optical distance \tV\ = 1 along a direction
slanted to the surface by angle $\alpha$ (figure \ref{fig:CG}). The optical
depth of the corresponding skin layer along the normal to the surface $\hat{\bf
n}$ is $\alpha$ at visual and $\alpha q_\nu$ at wavelength $\nu$, where $q_\nu
= \sigma_\nu/\sigma\sub{V}$ and $\sigma\sub{V}$ is the dust cross-section at
visual. In the case of a flat thin disk whose inner radius is determined by
dust sublimation, the grazing angle is
\eq{\label{eq:flat}
    \alpha\sub{flat} = {\alpha^\ast\over a},    \qquad \hbox{where}\quad
    \alpha^\ast = {4\over3\pi}{\R\over\Rs}, \quad a = {\ra\over\Rs};
}
here \R\ is the radius of the star, \Rs\ is the dust sublimation radius (eq.\
\ref{eq:R1}) and \ra\ is distance from the axis. Flaring is defined by the
radial profile of the disk height $H$ ($\ll$ \ra) or, equivalently,
\eq{\label{eq:beta}
    \beta = \arctan{H\over\ra} \simeq {H\over\ra}.
}
As is evident from figure \ref{fig:CG}, $\alpha = \gamma - \beta$ where
$\tan\gamma = dH/d\ra$, therefore the grazing angle of a flared disk is
\eq{\label{eq:alpha}
    \alpha = a{d\beta\over da}\,.
}
The CG surface layer serves as an effective optically thin disk atop the
underlying optically thick disk core, and its flux is obtained from the volume
integration listed in equation \ref{eq:flux}. In the case of face-on
orientation, the optical depth of the CG layer obeys $\sigma_\nu\!\int
n\sub{d}dz = q_\nu\alpha$, therefore its flux at distance $D$ is
\eq{\label{eq:FCG}
    F\sub{CG,\nu} = {2\pi\Rss\over D^2}\,q_\nu\!\!\int B_\nu(T)\,\alpha\,ada.
}
Since the temperature profile of optically thin dust depends only on distance
from the radiation source, the geometry dependence of this expression enters
only from the radial variation of the grazing angle $\alpha$ (reflecting the
dust column variation). But other geometries can produce an identical
expression. For example, in the case of spherical geometry the flux is
controlled by the dimensionless density profile $\eta$ (equation \ref{eq:eta})
and the radial optical depth is $\tau_\nu = \sigma_\nu\!\int\n dr$. Denoting
the optical depth at visual wavelengths by \tV, the flux is
\eq{\label{eq:Fsph}
  F\sub{sph,\nu} = {4\pi\Rss\over D^2}\,
    q_\nu\tV\!\!\int B_\nu(T)\,\eta\,y^2dy,
}
where $y = r/\Rs$. Since $y$ and $a$ enter only as integration variables in the
last two integrals, the two expressions are mathematically identical if
\eq{\label{eq:equiv1}
    \eta \propto {\alpha(y)\over y} \quad \hbox{and}\quad
    \tV = \half\int\!\! \alpha{da\over a}.
}
A minor point is that $T$ in eq.\ \ref{eq:FCG} is strictly a function of $y =
a\sqrt{1 + \beta^2}$ rather than $a$ (since temperature is controlled by
distance from the star); this slight difference can be ignored because $\beta
\ll 1$ everywhere in the disk.

Scattering can be treated similarly and produces the same result since the only
difference is that $B_\nu(T)$ is replaced by $J_\nu = L_\nu/4\pi r^2$, another
function of $y$ (see equation \ref{eq:scat}). Therefore, {\em there is a
complete equivalence between spherical halos and the CG surface layers of
flared disks}; either case defines a model with the other geometry and the
exact same flux. Specifically, equations \ref{eq:flat} and \ref{eq:equiv1} show
that a thin flat disk is equivalent to a spherical halo with $\eta \propto
1/y^2$ and \tV\ = \half$\alpha^\ast$. In the case of a flared disk, equations
\ref{eq:alpha} and \ref{eq:equiv1} show that the CG layer will produce
precisely the same flux as a spherical halo with
\eq{\label{eq:equiv}
    \eta \propto {d\beta \over dy},  \qquad
    \tV = \half\big[\beta(\Rd) - \beta(\Rs)\big];
}
in particular, the equivalent halo of a disk with flaring angle $\beta \propto
1/a^p$ has density distribution $\eta \propto 1/y^{p + 1}$. And
conversely---given a spherical halo, the flared disk with the same outer radius
and
\eq{\label{eq:beta2}
    \beta(a) = \beta(1) + 2\tV\int_1^a\eta(y)dy
}
will produce the exact same flux from its CG-surface layer.

Each disk defines a mathematically equivalent halo. Although this equivalence
was derived only for face-on viewing of the disk, it carries to most
inclination angles since the flux from optically thin dust involves a volume
integration (equation \ref{eq:flux}) and the observed fraction of the disk
surface layer remains largely intact as long as internal occultation is not too
significant. Similarly, the general analysis presented in appendix A shows that
the spherical idealization is not essential for the halo geometry. The dust
distribution can be flattened and even distorted into irregular shape before
severely affecting the emerging flux.

These results resolve the paradox of successful SED fits for the same star, GM
Aur, with a flared disk both with a halo (Schneider et al 2003) and without one
(Chiang \& Goldreich 1999). The halo contribution to the flux can be absorbed
into the disk component by readjusting the flaring law, enabling a successful
SED fit without a halo even though it is directly visible in imaging
observations. The problems Stapelfeldt et al (2003) report with disk-only
models of HV Tau C find a similar explanation. The halo contribution to the SED
can be shifted to the disk, with the halo optical depth added to the flaring
angle (equation \ref{eq:equiv}).  This leads to unreasonably large flaring, as
Stapelfeldt et al find. The same problem was encountered by Kikuchi et al
(2002) in modeling flat-spectrum TTS. The halo optical depth in these sources
are sufficiently high that they cannot be substituted by disks with realistic
flaring.

In addition to explaining the shortcomings of disk-only models in these
specific cases, the equivalence between halos and CG layers has two important
broad consequences:

\begin{enumerate}
\item
{\em When the disk is imbedded in a halo that radiates more than its CG layer,
the halo becomes the dominant component of the IR flux}. This happens when the
halo contains more dust than the halo-equivalent of the disk. From equations
\ref{eq:beta} and \ref{eq:equiv}, the IR radiation from the system is dominated
by the halo contribution whenever
\eq{\label{eq:equiv2}
    \tau\sub{halo} > \frac14{H(\Rd)\over\Rd}
}
where $\tau\sub{halo}$ is the optical depth across the halo at visual
wavelengths (2\tV\ for spherical halos).
\item
{\em It is impossible to distinguish the CG layer of a flared disk from a halo
with flux measurements}. Only imaging can produce an unambiguous signature of
the CG layer.
\end{enumerate}

\begin{figure}
\centering \leavevmode
 \psfig{file=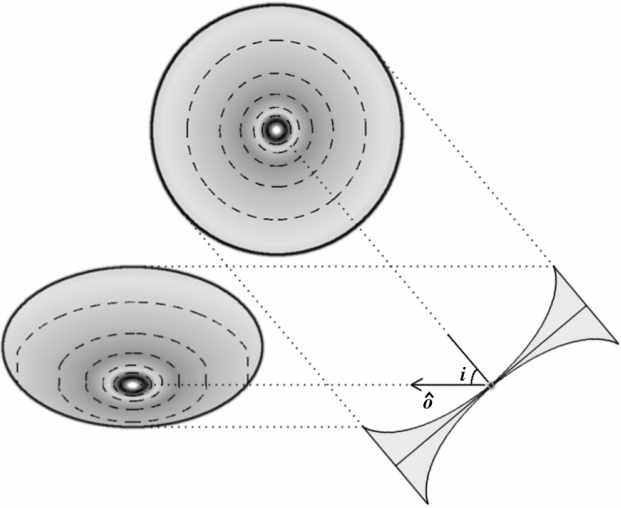,width=\figsize,clip}
\caption{Points on the surface of a flared disk at equal distance from the star
lie on a circle centered on the disk axis. The circle retains its shape in
pole-on viewing but is deformed into an off-center ellipse (equation
\ref{eq:contour}) in viewing from inclination angle $i$.} \label{fig:CGimage}
\end{figure}

\subsection{Images}

The brightness contours of face-on flared disks are concentric circles centered
on the star. Inclined viewing changes the contours substantially. Consider the
intensity of radiation scattered from the CG surface layer. It obeys
\eq{
        I \propto \frac{\tau(\hat{o})}{r^2}
}
where $r$ is the distance to the star and $\tau(\hat{o})$ the optical depth
toward the observer at the scattering point (equation \ref{eq:Basic}). Both
factors introduce distinct image asymmetries.

The fundamental reason for image distortion by inclination is that the same
projected distance from the star corresponds to widely different locations on
the surface of the disk. On that surface, contours of equal distance from the
star are circles of radius \ra. When viewed face-on from distance $D$, each
contour appears as a concentric circle of radius $\ta = \ra/D$, as seen in the
top image in figure \ref{fig:CGimage}. At inclination viewing angle $i$ to the
disk axis, the contour is no longer circular. Absent flaring, the contour
becomes an ellipse centered on the star with major axis 2\ta\ and minor axis
$2\ta\cos i$, aligned with the projection of the disk axis on the plane of the
sky. Flaring raises the contour to height $H = \ra\tan\beta$ above the
equatorial plane (figure \ref{fig:CG}), and the star is shifted toward the
observer along the minor axis by $\ta\tan\beta\sin i$. A point on the contour
at position angle $\phi$ from the near side of the minor axis is observed at
displacement $\theta = \ta g(\phi)$ from the star, where
\eq{\label{eq:contour}
 g(\phi) =
 \left[\left(\tan\beta\sin i - \cos i\cos\phi\right)^2 +
       \sin^2\phi\right]^{1/2}.
}
These contours are shown in the bottom image of figure \ref{fig:CGimage}. The
off-center position of the star on the minor axis creates an asymmetry such
that the far and near portions of this axis obey
$\theta\sub{far}/\theta\sub{near} = \cos(i - \beta)/\cos(i + \beta)$. Because
$\beta$ increases with \ta, this asymmetry increases with distance from the
star.

\begin{figure}
\centering \leavevmode
 \psfig{file=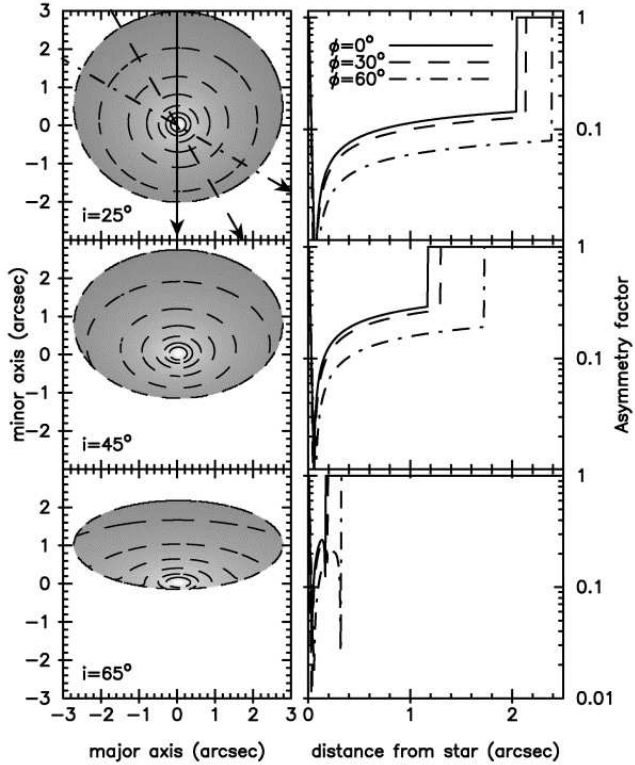,width=\figsize,clip}
\caption{Viewing at inclination angles $i$ = 25\deg, 45\deg\ and 65\deg\ of
radiation scattered off the surface of a flared disk with 3\arcsec\ radius and
constant grazing angle $\alpha$ = 4.9\deg. For each $i$, the left panel shows
the image with brightness contours, the right panel the asymmetry factor $A$
(equation \ref{eq:A}) along three azimuthal directions, shown in the top left
panel, whose angles are designated from the observer's direction. The asymmetry
parameter vanishes along the major axis ($\phi$ = 90\deg) at all inclination
angles and along any axis at $i$ = 0\deg\ and 90\deg. For flat disks, $A$ is
identically zero at all inclination angles.} \label{fig:asymmetry}
\end{figure}

At observed displacement $(\theta,\phi)$ from the star, a point on the surface
of the disk is located at $r \simeq \ra = D\theta/g(\phi)$. At that point the
optical depth of the CG layer toward the observer is $\tau(\hat{o}) =
q_\nu\alpha/o(\phi)$, where
\eq{
 o(\phi) = \hat{n}\cdot\hat{o} = \cos i\cos\gamma - \sin i\sin\gamma\cos\phi.
}
Therefore the scattering image obeys
\eq{\label{eq:CGsca}
 I(\theta,\phi) \propto \left({g(\phi)\over\theta}\right)^{\!2}
  {\alpha\over o(\phi)}.
}
In this expression, both $\alpha$ and $\gamma$ are determined at the location
$\ra = D\theta/g(\phi)$ on the disk surface. A power-law grazing angle $\alpha
\propto 1/r_{\rm a}^p$ produces the image $I(\theta,\phi) \propto
[g(\phi)/\theta]^{2 + p}/o(\phi)$. This expression and the resulting brightness
contours explain the results of Monte Carlo scattering calculations for flared
disks (Whitney \& Hartmann 1992, Wood et al 1998). Figure \ref{fig:asymmetry}
shows the scattering images at three viewing angles of a flared disk with the
parameters suggested for AB Aur by Dullemond et al (2001).

The images produced at emission wavelengths are handled in complete analogy.
The only change is the replacement of $r^{-2}$ by the temperature $T$, i.e.,
another function of $r$, modifying the dependence of brightness on
$g(\phi)/\theta$.

\subsubsection{Image asymmetry}
Brightness contours not subject to rim occultation are ellipses with
eccentricity $e = \cos i$ that directly determines the inclination angle
irrespective of the flaring profile. The images shown in figure
\ref{fig:asymmetry} possess an additional deviation from circular symmetry,
unique to flaring and conveniently measured by the brightness at diametric
locations across an axis through the star
\eq{\label{eq:A}
  A(\theta,\phi) =  {I(\theta,\phi + \pi) - I(\theta,\phi)\over
                     I(\theta,\phi + \pi) + I(\theta,\phi)}.
}
This asymmetry parameter vanishes for flat disks at all inclination angles, and
for pole-on and edge-on viewing irrespective of the flaring. However, at
intermediate inclination angles, flaring introduces substantial asymmetry, as
is evident from figure \ref{fig:asymmetry}.

Non-vanishing $A$ is the hallmark of inclined flared disks because it measures
the displacement of the isophote centers from the peak brightness position. Its
systematic variation with azimuthal angle easily distinguishes it from
deviations from the perfect geometry of idealized models or noise in the data.
Each flaring profile produces its own characteristic signature $A$. For
example, it is easy to show that the constant grazing angle $\alpha$ used in
figure \ref{fig:asymmetry} gives $A \simeq \tan\beta\tan i$ along the minor
axis. Therefore, measuring $A$ determines the flaring profile once the
inclination is determined from the eccentricity of the brightness contours.

\section{Halo-imbedded-Disks}

\begin{figure}
\centering \leavevmode
 \psfig{file=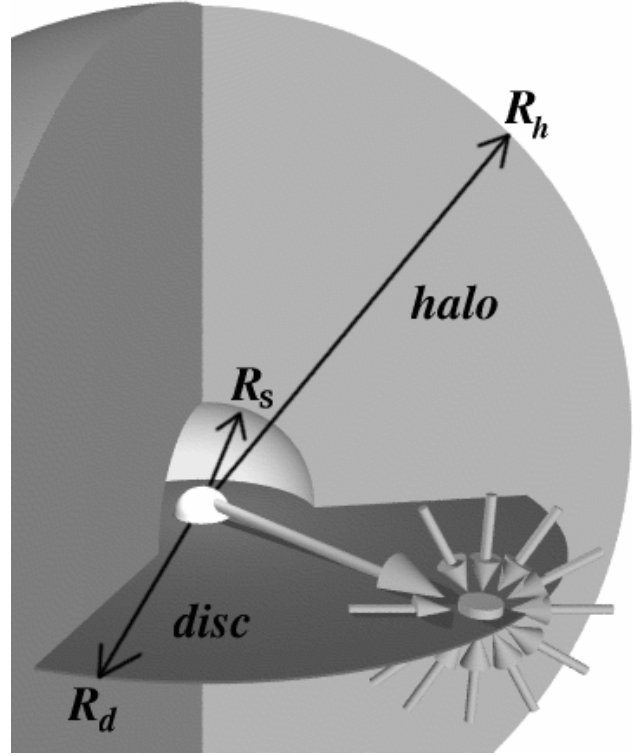,width=\figsize,clip}
\caption{Geometry of the halo-imbedded-disk model: a flat geometrically-thin
optically-thick disk extends from the stellar surface to radius \Rd. An
optically thin spherical halo extends from the dust sublimation radius \Rs\ to
\Rh. The small pillbox at the disk surface serves as a Gaussian surface for
flux conservation.} \label{fig:cartoon}
\end{figure}

The results of the previous section show that flux calculations of a flared
disk in vacuum can be considered a special case of the more general family of
models in which the disk is imbedded in an optically thin halo. In the CG case
the ``halo'' is the disk surface layer, fully determined from the flaring
geometry. This layer provides the same IR emission and heating of the
underlying optically thick core as its equivalent halo. Therefore, a study of
the general halo-imbedded-disk problem contains every possible CG model of
flared disks while also covering all cases in which the disk is indeed imbedded
in a dusty halo whose optical depth exceeds the bound in equation
\ref{eq:equiv2}.

Consider a star surrounded by a geometrically thin passive disk and a dusty
halo (figure \ref{fig:cartoon}).  We study the case of a flat disk and a
spherical halo, simplifications that enable us to derive analytical results and
broad conclusions that offer important insight. These simplifications do not
cause any serious limitations. As noted previously, the halo geometry can be
distorted considerably without much impact on the outcome. And because only the
optically thick core of the disk enters into considerations, its only relevant
property is its temperature distribution, the surface shape is immaterial.
Indeed, Wolf et al (2003) present detailed model calculation of a flared disk
imbedded in an envelope pinched around the equatorial plane, and their results
fully conform to our conclusions.

The halo extends from the inner radius \Rs\ to some outer radius $\Rh = Y\Rs$.
Thanks to scaling, instead of these radii we can specify the dust temperature
on each boundary (see appendix A). The halo is fully characterized by its
density profile $\eta(y)$ (eq.\ \ref{eq:eta}) and optical depth \tV; only $\tV
\la 1$ is relevant in TTS and Haebes since the star is always visible. Because
of its potentially large optical depth, the disk can extend inside the
dust-free cavity where its optical depth comes from the gaseous component. The
geometrically-thin disk assumption implies that the disk temperature varies
only with radius, vertical temperature structure is ignored. This temperature
is calculated from radiative flux conservation through the Gaussian surface in
the shape of a small pillbox shown in figure \ref{fig:cartoon}. Denote by \H\
the radiative flux entering the pillbox from above, including both the stellar
and diffuse components. Then
\eq{
 \sigma T^4 - 2\pi\int\!\!B_\nu(T)E_3(\tau^{\rm D}_\nu) d\nu
 + \int\limits_{2\pi}\!\!\mu I_\nu e^{-\tau^{\rm D}_\nu/\mu}d\Omega d\nu
 = \H,
}
where $E_3$ is the 3rd exponential integral, $\tau^{\rm D}_\nu$ is the disk
vertical optical depth and $\mu$ is the cosine of the angle from the disk
normal. The first two term on the left are the disk contribution to the upward
flux, the third is the contribution of local intensity transmitted upward
through the disk. When the disk is optically thick at frequencies around the
peak of $B_\nu(T)$ ($\tau^{\rm D}_\nu \gg 1$ for $\nu \sim kT/h$), the second
and third terms can be neglected, leading to the standard expression for disk
temperature (e.g.\ Friedjung 1985, Kenyon \& Hartmann 1987). We assume this to
be the case everywhere in the disk, an assumption that we check for
self-consistency in all our model calculations.

With its temperature derived, the disk emission is calculated from $B_\nu(T)[1
- \exp(-\tau^{\rm D}_\nu/\mu)]$. Because the disk is optically thick around the
Planckian peak at all radii, the self-absorption factor can be neglected in the
calculation of the disk overall bolometric flux. Then $\Fdisk(D,i)$, the disk
flux observed at distance $D$ and inclination $i$, is proportional to $\cos i$,
reflecting the variation of projected area. This proportionality remains
largely unaffected by the envelope attenuation because the short wavelengths,
the main contributors to the bolometric flux, emanate from close to the star so
that their pathlength is approximately isotropic. Denote by \Ldisk\ the disk
contribution to the overall luminosity $L$ and by \Lsph\ the contribution of
the (halo + attenuated stellar) spherical component.  The corresponding flux
components are then
\eq{\label{eq:fluxes}
    \Fdisk(D,i) = \frac{\Ldisk}{2\pi D^2}\,\cos i, \qquad
    \Fsph(D) = \frac{\Lsph}{4\pi D^2}
}
and the overall flux is
\[
    F(D,i) = \Fdisk(D,0)\cos i + \Fsph(D)
           = \frac{L}{4\pi D^2}\cdot\frac{1 + 2x\cos i}{1 + x}
\]
where $x = \Ldisk/\Lsph$. The standard ``bare'' disk has $\Ldisk = \frac14L$
(Kenyon \& Hartmann 1987), therefore in this case $x = \frac13$. Larger
fractions can occur when the disk is imbedded in a halo because of the heating
effect of the diffuse radiation, discussed below.

We performed detailed model calculations with the code DUSTY (Ivezi\'c, Nenkova
\& Elitzur 1999) which takes into account the energy exchange between the star,
halo and disk, including dust scattering, absorption and emission. Because its
optical depth is typically $\tV \la 1$, the halo is transparent to the disk
emission in all the models we consider and we neglect the disk effect on the
halo. In all the calculations, the spectral shapes $q_\nu$ of the grain
absorption and scattering coefficients are those of standard interstellar mix,
the sublimation temperature \Ts\ = 1500~K. The spectral shape of the stellar
radiation is taken from the appropriate Kurucz (1994) model atmosphere.

\subsection{Temperature Profiles}
\label{sec:Temp}
\begin{figure}
\centering \leavevmode
 \psfig{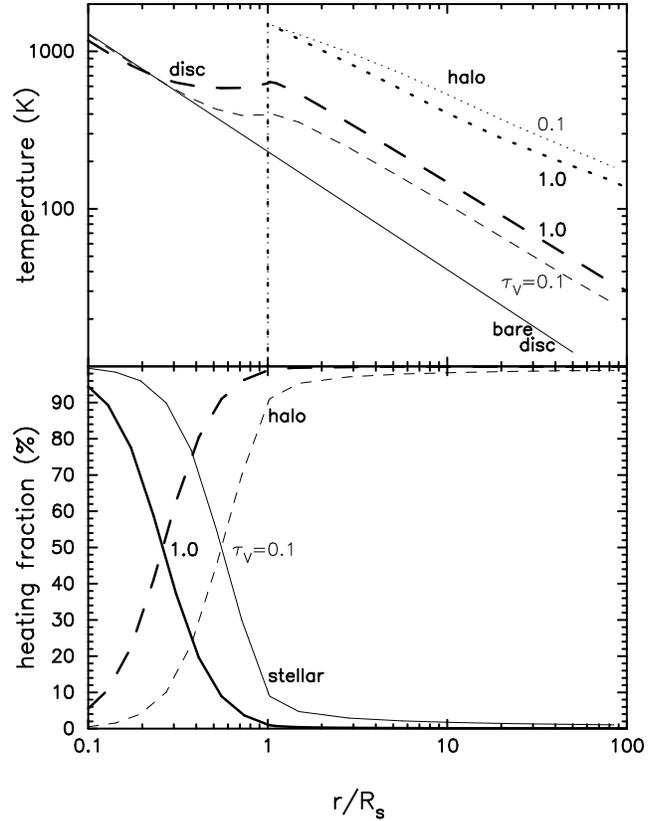}
\caption{Top: Temperature profiles of a disk when heated only by a central star
with \T\ = 10,000 K (full line), and when imbedded in a spherical dusty halo
with \tV\ = 0.1 or 1, as marked. The halo starts at dust sublimation \Ts\ =
1,500 K and its density profile is $\propto r^{-2}$. The temperature profile of
the halo is also shown in each case. Bottom: The fractional contributions of
the halo and (attenuated) stellar components to heating of the disk}
\label{fig:T}
\end{figure}
\begin{figure*}
\begin{minipage}{\textwidth}
\centering \leavevmode
 \psfig{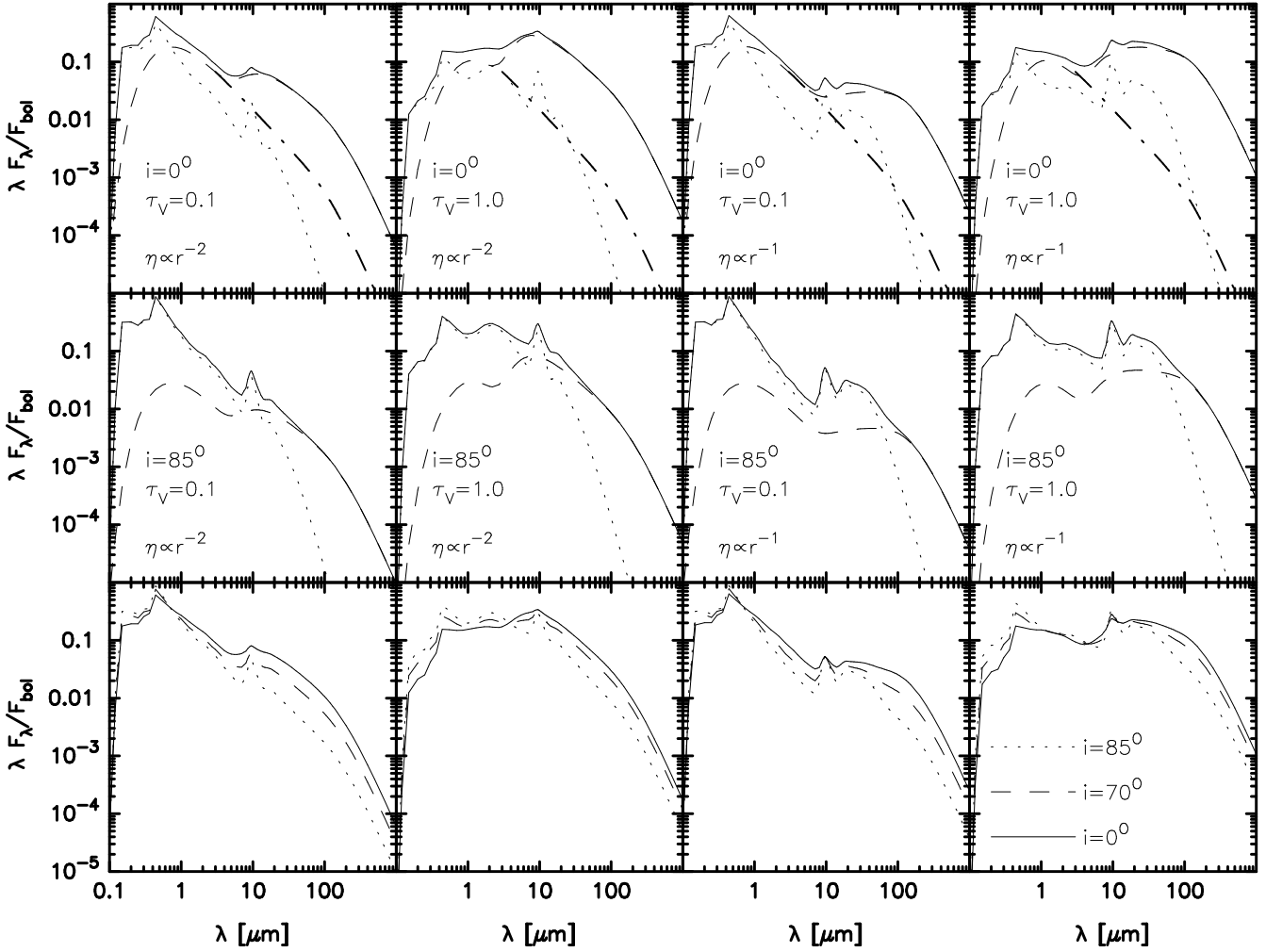}
\caption{Sample SEDs for halo-imbedded-disks around stars with \T\ = 10,000 K.
The halo starts at \Rs\ where \Ts\ = 1,500 K and extends to 1,000\Rs, with
density profile $\eta$ and optical depth \tV\ as indicated. The disk starts at
the stellar surface and extends to the radius \Rd\ set by the edge temperature
\Td\ = 25 K (see Table 1). Each SED is normalized with the bolometric flux
$F\sub{bol} = F(D,i)$ of the appropriate viewing angle (see equation
\ref{eq:fluxes}). Top row: The SED for pole-on viewing. Full lines denote the
overall flux, dotted lines the spherical (halo + attenuated stellar) component
and dashed lines the disk component. The thick dot-dashed line is the flux from
a face-on ``bare'' disk in the absence of an imbedding halo. It is omitted in
the other rows. Mid row: Same models viewed at inclination angle $i$ = 85\deg.
Bottom row: Variation of the overall SED with viewing angle $i$. Results for $i
<$ 70\deg\ are barely distinguishable from $i$ = 0\deg.
} \label{fig:SEDs}
\end{minipage}
\end{figure*}

The heating rate of a thin flat disk by the stellar radiation at $r \gg \R$ is
\eq{\label{eq:Hstar}
   \H_\ast = {2\Fs\over3\pi}{\R\over\Rs} {1\over a^3},
}
where \Fs\ is the stellar flux at \Rs\ and $a = \ra/\Rs$, with \ra\ distance
from the axis (Friedjung 1985). This result reflects the $1/a^2$ decline of the
stellar solid angle and the $1/a$ variation of the grazing angle, yielding disk
temperature variation $T \propto a^{-3/4}$. Natta (1993) noted that imbedding
the disk in a dusty halo can significantly affect its temperature even at small
halo optical depths (see also D'Alessio, Calvet \& Hartmann 1997 for the effect
of optically thick halos). With a simple model for scattering at a single
wavelength she found that the disk temperature law becomes $T \propto a^{-(1 +
p)/4}$ if the halo density profile is $\eta \propto y^{-p}$.

Our calculations confirm this important point. Figure \ref{fig:T} shows the
temperature profile for a disk around a \T\ = 10,000 K star when ``bare'' and
when imbedded in a spherical halo with $\eta \propto y^{-2}$ and \tV\ = 0.1 and
1\footnote{The addition of a halo can only raise the disk temperature, yet
figure \ref{fig:T} shows that our calculated profile for \tV = 1 dips slightly
below that of the bare disk at $a \la\ 0.2$. This happens because we neglect
the disk emission in the calculation of the halo temperature. The error
introduced by this approximation is of order 2\% when \tV = 1, and even less at
smaller \tV.}. Even though a halo with \tV\ as small as 0.1 is almost
transparent to the stellar radiation, it still causes a large rise in disk
temperature. As is evident from the bottom panel, the halo contribution to the
disk heating overtakes the stellar contribution inside the dust-free cavity and
dominates completely once the dust is entered. Wolf et al (2003) present a
similar figure for their model.

A dusty envelope with \tV\ = 0.1 intercepts only about 10\% of the stellar
luminosity while the disk intercepts 25\% of that luminosity. So how can the
halo dominate the disk heating? The reason is that direct heating of the disk
by the star occurs predominantly at small radii. The disk absorbs more than
90\% of its full stellar allotment within $10\R$ while its entire remaining
area, even though so much larger, absorbs only $0.025L$. From equation
\ref{eq:R1}, the halo starts at $\Rs \sim 100\R$, where $\H_\ast$ has already
declined to \about\ \E{-6} of its value at the stellar surface. In contrast,
the halo emission is isotropic, therefore half of the radiation it intercepts
is re-radiated toward the disk, greatly exceeding the direct stellar
contribution. Further insight can be gained from the approximate solution
presented in Ivezi\'c \& Elitzur (1997; IE hereafter) for radiative transfer in
spherical symmetry. From equations 20 and B4 of IE it follows that disk heating
by a halo with $\tV \la 1$ and $\eta \propto y^{-p}$ is roughly
\eq{
     \H\sub{h} = {3\Fs\over8}{p - 1 \over p + 1}\,\tV\,
             \x\cases{1                        &\qquad $a < 1$  \cr\cr
                     {1/a^{1 + p}}  &\qquad $a > 1$  }
}
when $p > 1$; when $p = 1$ the factor $(p-1)/(p+1)$ is replaced by $1/(2\ln
Y)$. This yields $T \propto a^{-(1 + p)/4}$, corroborating Natta's result and
extending its validity beyond the single-wavelength scattering approximation
she employed. The temperature profile is similar to that of a bare disk when $p
= 2$ but more moderate when $p < 2$. More importantly,
\eq{
 {\H\sub{h}\over \H_\ast}\bigg|_{a = 1}\!\!\!=
                {9\pi\over16}{p-1\over p+1}\,{\Rs\over\R}\,\tV.
}
And since \Rs\ \about\ 100\R, the halo dominates the heating at $a = 1$ for
\tV\ as small as 0.02 when $p = 2$. As \tV\ increases, the halo dominance of
the heating moves inside the cavity. There the halo heating remains
approximately constant while the stellar heating varies as $a^{-3}$, therefore
stellar heating dominates only at $a \la (60\tV)^{-1/3}$, at larger distances
the halo takes over. This explains the results presented in the lower panel of
figure \ref{fig:T} as well as figure 9 in Wolf et al (2003).

The figure also shows the temperature profile of the halo. This profile is
largely independent of \tV, varying roughly as $y^{-2/(4 + n)}$ when the
long-wavelength  spectral shape of the dust absorption coefficient is $q_\nu
\propto \nu^n$. The important property evident in the figure is that the disk
is much cooler than the envelope at all radii at which both exist and can also
contain cooler material in spite of being much smaller, with far reaching
consequences for the system radiation.

\subsection{SED}

From equation \ref{eq:fluxes}, the fractional contribution of the disk to the
overall bolometric flux is
\eq{\label{eq:rho}
    \rho = {\Fdisk\over \Fdisk + \Fsph} = {2x\cos i \over 1 + 2x\cos i}\,.
}
Face-on orientation gives the maximal $\rho = 2x/(1 + 2x)$ and the standard
``bare'' disk, with $x = \frac13$, has $\rho \le \frac25$.  Introduce the
normalized SED $f_\nu = F_\nu/\!\int\!F_\nu d\nu$, with similar, separate
definitions for the disk and spherical components of the flux. Then
\eq{\label{eq:f}
    f_\nu = \rho f\sub{\nu,disk} + (1 - \rho)f\sub{\nu,sph}
}
Since the disk flux obeys $F_{\nu,\rm disk}(i) = F_{\nu,\rm disk}(0)\x\cos i$
for the range of parameters considered here, $f\sub{\nu,disk}$ is independent
of the viewing angle $i$, and the entire $i$-dependence of the SED comes from
the mixing factor $\rho$.

Figure \ref{fig:SEDs} shows sample SEDs for some representative models. The
halos extend from \Rs\ to 1,000\Rs, with density profiles and overall optical
depths as indicated. The behavior of SEDs for spherical shells was discussed in
IE and since the halo emission is unaffected by the imbedded disk, the SEDs
plotted in dashed lines need no further discussion. The disk, on the other
hand, is strongly affected by the halo as is evident from contrasting each disk
SED, plotted in long-dashed line, with what it would have been in the absence
of a halo (dot-dashed line).  The two curves are identical within the first
bump around 1 \mic, caused by the stellar heating. In the absence of a halo,
the disk SED drops from that peak as $\lambda F_\lambda \propto
\lambda^{-4/3}$. However, halo heating of the outer regions of the disk
generates the second broad bump of disk radiation, which is almost two orders
of magnitude higher than the ``bare'' disk emission.

The halo heating effect is also evident from other disk properties. The disks
in these models start at the stellar surface and extend to a radius \Rd\ where
the temperature is \Td\ = 25 K. In the absence of a halo, this temperature
would be reached at \Yd\ = \Rd/\Rs\ = 18.  As Table 1 shows, heating by even a
tenuous halo with \tV\ = 0.1 pushes this radius out by almost factor 5 for the
steep density profile $r^{-2}$ and another factor of 2 for the flatter $r^{-1}$
profile which spreads the heating further away from the star. The impact of
halo heating increases with \tV, pushing \Rd\ further out still. Similarly, the
disk luminosity, $xL/(1 + x)$, is only $0.25L$ in the absence of a halo but
increases as the halo directs more radiation toward it to the point that it
becomes $0.6L$ when \tV\ = 1.

\def\d{\phantom{1}}
\begin{table}
\begin{center}
\begin{tabular}{ccccc}
           & \multicolumn{2}{c}{\tV = 0.1}
           & \multicolumn{2}{c}{\tV = 1}   \\ \cline{2-5}
           & x    &  \Yd &   x  & \Yd      \\ \hline
  $r^{-1}$ & 0.43 &  190 & 1.33 & 400      \\
  $r^{-2}$ & 0.47 & \d85 & 1.56 & 135      \\ \hline
\end{tabular}
\end{center}
\caption{Derived parameters for the models whose SEDs are presented in figure
\ref{fig:SEDs}. The luminosity ratio of the components is $x = \Ldisk/\Lsph$
and the disk radius is \Rd\ = \Yd\Rs, set from the requirement \Td\ = 25 K. A
``bare'' disk (\tV\ = 0) has $x = \frac13$ and \Yd\ = 18.
}
\end{table}

\begin{figure}
\centering \leavevmode
 \psfig{file=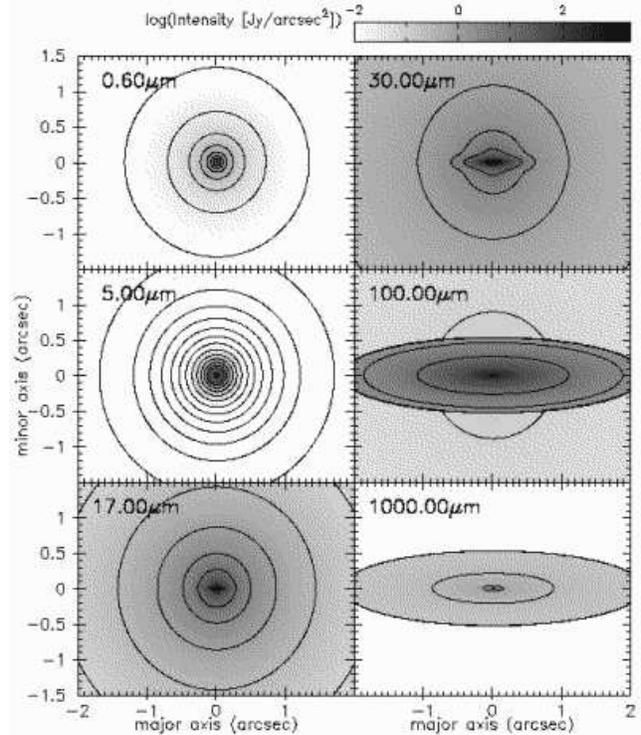,width=\figsize,clip}
\caption{Images at various wavelengths of a halo-imbedded-disk around a star
with \T\ = 10,000 K. The halo starts at \Rs\ where \Ts\ = 1,500 K and extends
to 1,000\Rs, with density profile $\eta \propto y^{-2} + 0.05y^{-1}$ and \tV\ =
0.6. The disk starts at the stellar surface and extends to radius \Rd\ set by
\Td\ = 25 K. The angular scale corresponds to bolometric flux $F\sub{bol}$ =
\E{-10} W m$^{-2}$. The viewing inclination angle is $i$ = 76\deg.}
\label{fig:Images}
\end{figure}

Although it is more compact, the disk can become the stronger emitter at long
wavelengths so that the overall SED is dominated by the halo at IR wavelengths
and by the disk at sub-mm and mm wavelengths. This role reversal affects also
the wavelength behavior of images. Figure \ref{fig:Images} shows a series of
images at various wavelengths for a sample model. At IR wavelengths the image
is dominated by the halo, displaying the size variation discussed in appendix
A. The 0.6\mic\ image is dominated by scattering, the $\lambda \ge$ 5\mic\
images reflect dust emission, leading to size increase with wavelength. The
disk emerges at 17\mic\ and dominates the $\lambda \ge$ 100\mic\ images. The
finite beam size and dynamic range of any given telescope could result in an
apparent size decrease between 10 \mic\ and 100 \mic\ in this case. Such an
effect has indeed been discovered in the Haebes MWC 137, whose observed size
decreases between 50 \mic\ and 100 \mic\ (DiFrancesco et al 1994, 1998). A
switch from envelope to disk domination provides a simple explanation for this
puzzling behavior. No single dust configuration can explain such a decrease, a
conclusion reached already in MIVE and further affirmed by the results of
appendix A. DiFrancesco et al (1998) suggest that this behavior might reflect
multiple, rather than singular, sources of heating but the results of appendix
A show the inadequacy of this conjecture. The region heated by any single
source displays an increase of observed size with wavelength, and the
superposition of multiple heating sources preserves this behavior. The opposite
trend is possible only when the density distribution contains two distinct
components, one optically thick, cool and compact, the other optically thin,
warmer and more extended.

A similar effect was detected also in the dust-shrouded main-sequence star
Vega. Van der Bliek, Prusti \& Waters (1994) find that its 60 \mic\ size is
$35\arcsec \pm 5\arcsec$, yet 850 \mic\ imaging by Holland et al.\ (1998)
produced a size of only $24\x21\arcsec \pm 3\arcsec$. So the dust distribution
around Vega, too, could contain both spherical and disk components with the
switch from halo- to disk-dominance occurring somewhere between 60 and 850
\mic. Indeed, imaging at 1.3mm by Willner et al (2002) reveal the presence of
the disk.

\begin{figure}
\centering \leavevmode
 \psfig{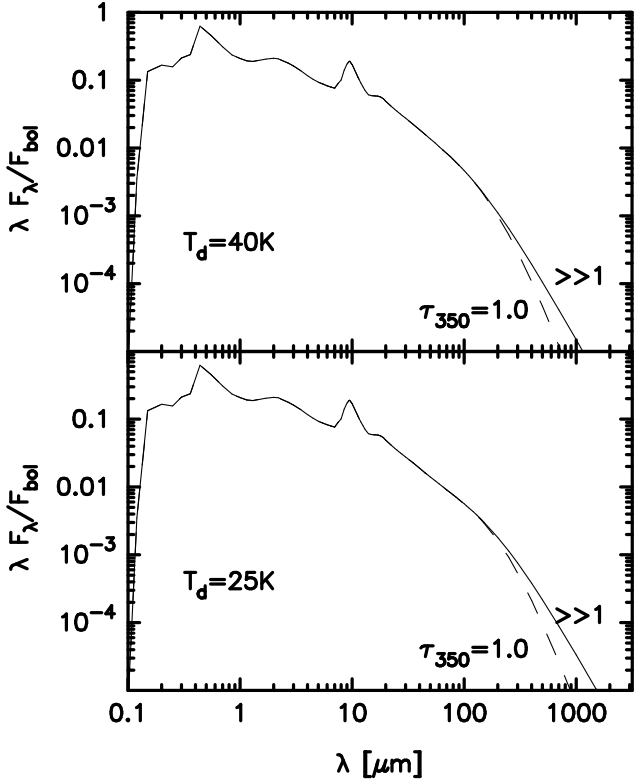}
\caption{The effect on the SED of varying the temperature (\Td) and 350\mic\
optical depth ($\tau_{350}$) of the disk outer edge. The displayed model has
\tV\ = 0.5 and $\eta \propto r^{-2}$. The viewing angle is 85\deg. Dashed lines
correspond to $\tau_{350}$ = 1, full lines to $\tau_{350}$ sufficiently large
that the disk edge remains optically thick at all displayed wavelengths.}
\label{fig:Td}
\end{figure}

The calculation of the disk emission contains two free parameters, involving
the temperature and optical depth at the disk edge. The disk outer radius \Rd\
determines its lowest temperature \Td\ and a corresponding Planck-peak
wavelength. While shorter wavelengths are emitted from a range of disk radii,
all longer wavelengths originate from the edge of the disk. The resulting
effect can be seen in figure \ref{fig:Td}, which shows two representative
values of \Td. By the model assumptions the disk must be optically thick at the
local Planck-peak wavelength everywhere, \about\ 350 \mic\ at \Td\ = 25 K. As
long as the disk edge remains optically thick also at longer wavelengths, the
emission follows the Rayleigh-Jeans profile $f\sub{\nu,disk} \propto \nu^2$.
Once the disk edge becomes optically thin, the SED switches to the steeper
decline $f\sub{\nu,disk} \propto \nu^2\sigma_\nu$ at longer wavelengths. The
break in the disk SED is controlled by the optical depth of the disk edge,
which we specify at 350 \mic. Figure \ref{fig:Td} shows also the effects of
varying $\tau_{350}$ from its smallest value $\tau_{350} \sim 1$ to a value
sufficiently large that the edge is optically thick at all the displayed
wavelengths.

\begin{figure}
\centering \leavevmode
 \psfig{file=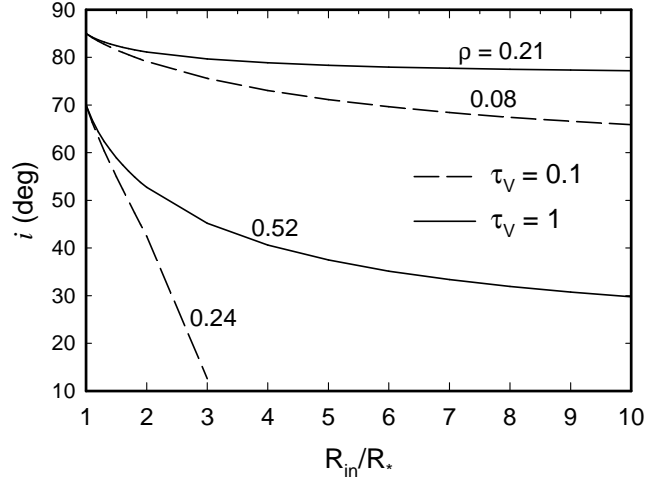,width=\figsize,clip}
\caption{Contours of fixed mixing coefficient $\rho$ (equation \ref{eq:rho}),
as marked: The SEDs are the same as those presented in figure \ref{fig:SEDs}
for $i$ = 70\deg\ and 85\deg\ when the disk inclination $i$ and its inner
radius \Rin\ vary together along each of the plotted curves. The contours are
virtually the same for the $r^{-1}$ and $r^{-2}$ halo density profiles.}
\label{fig:inclinations}
\end{figure}

\subsubsection{The disk inner radius}

The bottom panel of each model in figure \ref{fig:SEDs} shows the variation of
the overall SED with viewing angle $i$. The entire variation comes from the
mixing factor $\rho$ (see equations \ref{eq:rho} and \ref{eq:f}). Since the
parameters $x$ and $i$ enter only in the product $x\cos i$ but not separately,
the SED is subject to a degeneracy: systems viewed at different inclination
angles will have the same SED if they have the same $x\cos i$ in addition to
all other properties. Because of the rapid decline with distance of the
radiation absorbed from the star (cf eq.\ \ref{eq:Hstar}), the disk luminosity,
i.e., $x$, has a steep dependence on its inner radius \Rin; moving the disk
inner edge from \R\ to only 2\R\ removes 56\% of the stellar luminosity
intercepted by the disk, 3\R\ results in a 72\% removal. {\em Such central
holes reduce $x$ but do not impact any other relevant property} because they
remove only the hottest disk material whose contribution to the overall flux is
negligible in comparison with the stellar component.

Figure \ref{fig:inclinations} plots contours in the $i$--\Rin\ plane of
constant mixing factor $\rho$. It shows, for example, that the SEDs presented
in figure \ref{fig:SEDs} for $i$ = 70\deg\ would be the same for systems viewed
at $i$ = 35\deg\ if the disk inner radius is increased from 1\R\ to 2.2\R\ in
the \tV\ = 0.1 case and 6\R\ for \tV\ = 1. Although the sizes of these holes
cannot be determined from SED modeling of single stars, from statistical
arguments MIVE conclude that their existence is essential to produce a
plausible distribution of inclination angles.

\subsubsection{Flared disks without halos}

Thanks to the CG layer--halo equivalence, each model presented here describes
also a disk with no halo and with the flaring angle defined by eq.\
\ref{eq:beta2} from $\eta$ and \tV. In particular, the SED presented for $\eta
\propto 1/y^2$ halos describe also flared disks without halo and with $\alpha
\propto 1/a$, those with $\eta \propto 1/y$ cover flared disks with constant
grazing angle $\alpha$ ($\beta \sim \ln\,a$). The equivalence strictly holds
only for disks and halos of the same size. However, the models presented would
be little affected if each halo was truncated at \Rd\ because that would only
remove halo emission at long wavelengths where the SED is dominated by the disk
anyhow. Since the flaring angle of each equivalent disk reaches 2\tV\ at its
outer edge, halos whose optical depths require excessive flaring cannot be
realized with disks only. This was the problem recognized by Stapelfeldt et al
(2003) and Kikuchi et al (2002).

The equivalence holds only for the SED. High-resolution observations would
produce widely different images for each halo and its equivalent flared disk,
except when the latter is viewed face-on.

\section{Discussion}

The results of \S\ref{sec:CG} show that every disk imbedded in a halo with $\tV
\la \half$ can be replaced by a flared disk without a halo and with an
identical flux. This mathematical degeneracy explains the success of SED
modeling with disk alone a system like GM Aur (Chiang \& Goldreich 1999) even
though the halo was subsequently discovered in imaging observations (Schneider
et al 2003).  It also explains why such modeling runs into difficulties and
requires excessive flaring when the halos have larger \tV, as is the case in HV
Tau C (Stapelfeldt et al 2003) and flat-spectrum TTS in general (Kikuchi et al
2002).

Disks generally do not exist in pure vacuum. Equation \ref{eq:equiv2} defines
the circumstances under which the surrounding dust becomes the dominant
component of the IR flux.  Even at smaller \tV, when not dominating the overall
flux, the halo can still dominate the disk heating and make a strong impact on
its temperature profile (\S\ref{sec:Temp}). Ignoring the surrounding material
can produce misleading results regarding the disk properties, such as its
flaring profile.

In spite of the attractiveness of the flared disk as a simple, physical model
without any additional components, imaging observations give irrefutable
evidence for the existence of extended halos in pre main-sequence stars. The
origin of these halos has not been studied yet. Stapelfeldt et al (2003)
suggest that a replenishment process, either continued infall from the
surrounding ISM or a dusty outflow from the source itself, is operating. It is
noteworthy in this regard that accretion with the small rates of \about\ \E{-8}
\Mo~yr$^{-1}$ has been deduced from UV spectra of both Haebes (Grady et al.
1996) and TTS (Valenti, Basri \& Johns 1993; Gullbring et al 1998; Lamzin,
Stempels \& Piskunov 2001), and is consistent with halos that have \tV\ \about\
0.1 (MIVE). These low rates cannot correspond to the main accretion buildup of
the star but rather a much later phase, involving small, residual accretion
from the environment. The corresponding accretion luminosities are only \about\
0.1 \Lo, justifying their neglect in our calculations.

In addition to the CG layer--halo equivalence, our results reveal numerous
degeneracies that underscore the severe limitations of attempts to determine
the dust morphology from SED analysis without imaging observations. The SED of
a halo-imbedded-disk remains the same when the viewing angle and the size of
the disk central hole vary together as shown in figure \ref{fig:inclinations}.
From the results of appendix A, the SED of a spherical shell with power-law
density profile $1/r^p$ displays a dependence on $p$ only when $p \la 2.6$, and
then only in the spectral region $\lambda \le \lout$. All other regions of $p$
and $\lambda$ produce the same universal behavior $F_\nu \propto
\nu^2\sigma_\nu$ (\S\ref{sec:Spherical}). The dust optical properties introduce
additional degeneracies. The results of \S\ref{sec:Spherical} show that the
frequency dependence of $\sigma_\nu$ and the radial dependence of the density
profile can be interchanged on occasion without affecting the SED. The
fundamental reason for all these degeneracies is that the flux from an
optically thin source involves a volume integration (eq.\ \ref{eq:flux}) that
tends to remove much of the dependence on the underlying morphology. Although
the specific degeneracies we uncovered involve spherical geometry, the general
analysis in appendix A shows that the spherical idealization is not essential.
The dust distribution can be flattened and even distorted into irregular shape
before severely affecting the results.

These degeneracies make it impossible to determine the geometry from a fit to
the SED alone without additional input. Only imaging can trace the actual
density distribution, and scattering provides a more faithful presentation
because, unlike emission which involves also the dust temperature, it involves
only the density distribution. Reliance on SED modeling alone can produce
misleading results, as was the case for the parameters of the flared disk in GM
Aur.

\section*{Acknowledgments}

The partial support of NSF and NASA is gratefully acknowledged. \v{Z}.I.
acknowledges generous support by Princeton University.

\let\Ref=\item
{}

\appendix
\section {Optically Thin Dust}
\label{appendix}

\begin{figure}
\centering \leavevmode
 \psfig{file=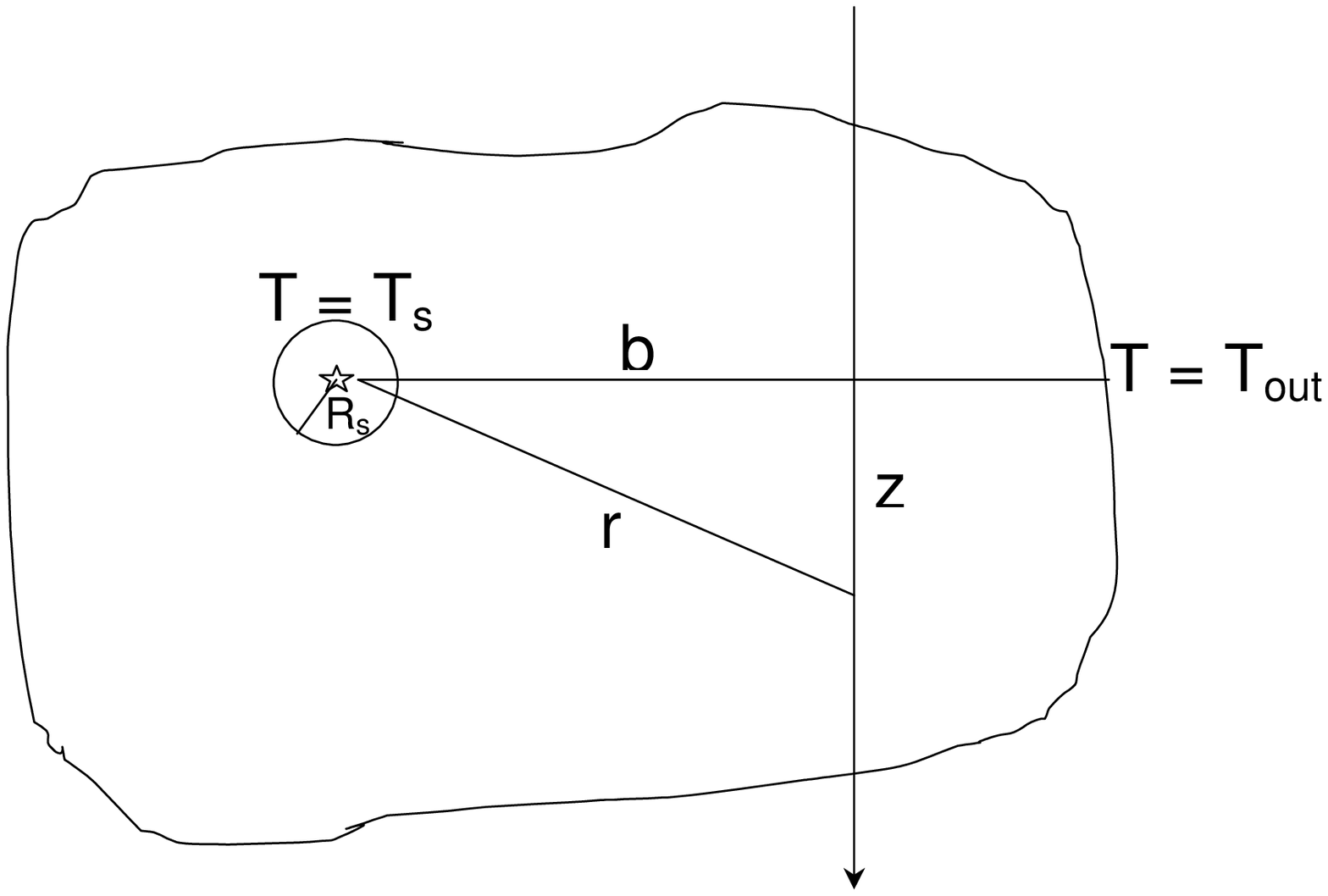,width=\figsize,clip}
\caption{A star imbedded in a cloud clears out a dust-free cavity of radius
\Rs, corresponding to dust sublimation temperature \Ts\ (equation \ref{eq:R1}).
The dust temperature declines with radial distance toward its surface value
\Tout. The intensity at impact parameter $b$ is obtained from integration along
the indicated path toward the observer.} \label{fig:imaging}
\end{figure}

Consider a cloud heated from inside by a star of radius \R\ and effective
temperature \T\ (figure \ref{fig:imaging}). The star clears out a dust-free
cavity of radius \Rs, determined by dust sublimation $T(\Rs) = \Ts$. When the
dust is optically thin, the cavity radius can be found from
\eq{\label{eq:R1}
  {\Rs\over\R} = \half\left(\bar{\sigma}(\T)\over\bar{\sigma}(\Ts)\right)^{1/2}
                    \left(\T\over\Ts\right)^2
}
where $\bar{\sigma}(T)$ is the Planck average at temperature $T$ of the
absorption cross section $\sigma_\nu$ (IE). With standard interstellar dust and
\Ts\ = 1500 K, the cavity radius obeys $\Rs/\R \simeq 100$ at a typical Haebes
temperature \T\ = 10,000 K.  In TTS, on the other hand, the dust is much closer
to the star: \Rs/\R\ is only 15 at \T\ = 5,000~K and as small as 3 at \T\ =
3,000 K. The intensity at frequency $\nu$ and impact parameter $b$ (fig.
\ref{fig:imaging}) is
\eq{\label{eq:Basic}
    I_\nu(b) = \sigma_\nu\int
    \left[(1 - \varpi_\nu)B_\nu + \varpi_\nu J_\nu \right]\,\n\,dz,
}
assuming isotropic scattering. Here \n\ is the dust density, $z$ is distance
along the path to the observer, $\varpi_\nu$ is the albedo at frequency $\nu$
and $J_\nu = \int I_\nu d\Omega/4\pi$. This expression neglects self-absorption
by the dust; the error in this approximation is of order $1 -
\exp\left(-\int\sigma_\nu\n dz\right)$.

\subsection{Scattering Wavelengths}

Since the dust temperature cannot exceed the sublimation temperature \Ts, there
is no dust emission at $\lambda \la$ 4\mic\x(1000K)/\Ts, only scattering.
Diffuse radiation and attenuation between the star and the scattering point can
be neglected since our discussion is centered on optically thin dust.  Then the
only source of scattering is the stellar radiation with energy density $J_\nu =
L_\nu/4\pi r^2$, where $L_\nu$ is the stellar luminosity at frequency $\nu$.
From equation \ref{eq:Basic}, the scattered brightness is
\eq{\label{eq:scat}
    I_\nu(\theta) = \frac{L_\nu}{4\pi }\,\varpi_\nu\sigma_\nu
               \int {\n\,dz\over r^2}\,,
}
where $\theta = b/D$ and $D$ is the distance to the observer. Since the
frequency- and geometry-dependence separate out, all scattering wavelengths
share a common image. Only the brightness level varies with $\nu$, and because
of the wavelength decline of $\varpi_\nu\sigma_\nu$ the observed size generally
{\em decreases with wavelength} when traced to the same brightness level. In
any geometry the scattering image always traces directly the variation of
column density along the line of sight, the dust temperature profile is
irrelevant.

\subsection{Emission Wavelengths}

At wavelengths longer than \about3\mic, $\varpi_\nu < \E{-2}$ and scattering
can be neglected. The Planckian enters in equation \ref{eq:Basic} as a function
of $T$ at fixed $\nu$, which can be well approximated by its Rayleigh-Jeans
limit at $T \ge T_\nu = 0.56h\nu/k$ and a sharp cutoff at $T_\nu$. With this
approximation the integration is limited to locations along the path where $T
\ge T_\nu$; regions with $T < T_\nu$ are too cold to emit appreciably at
frequency $\nu$. Excluding highly patchy geometries, the highest temperature on
the path occurs at $r = b$ (i.e., $z = 0$), the closet distance to the star,
and only paths with $T(b) > T_\nu$ contribute to the brightness. As $z$
increases in either direction, $T$ decreases. The integration is truncated
either because the temperature becomes too low, in which case the emission is
{\em temperature bounded}, or because the edge of the source is reached and the
emission is {\em matter bounded}. Denote the resulting integration limits $Z_i$
($i$ = 1, 2), then
\eq{
  I_\nu(\theta) = {2\over c^2}\,\nu^2 \sigma_\nu\!\!
             \int_{Z_1}^{Z_2}\!\! kT\,\n\, dz.
}
In the matter bounded case $Z_i$ is the edge of the source, the integral is
independent of $\nu$ and the frequency dependence of the intensity follows
$\nu^2\sigma_\nu$. In the case of temperature bounded emission the integration
limits introduce additional $\nu$-dependencies that modify this behavior.
However, the integration can be extended to $\infty$ whenever (1) $Z_i \gg b$
and (2) the product $\n T$ of dust density and temperature declines along the
path faster than $1/z$. Therefore, {\em when these two conditions are met, the
frequency variation of optically thin emission is $I_\nu \propto
\nu^2\sigma_\nu$ even when it is temperature bounded}. Independent of geometry,
all frequencies that obey these conditions produce a common image, similar to
the scattering case; only the scale of brightness varies with $\nu$. This
result makes it possible to determine the wavelength dependence of the dust
cross section directly from imaging observations.

Similar to the variation along the line of sight, when $b$ increases (moving
away from the star) the emission again is truncated by either the matter or
temperature distribution. Denote by \Tout\ the temperature at the source outer
edge. The corresponding emission cutoff wavelength is
\eq{
    \lout =  100\mic\x{40 \rm K \over \Tout}.
}
When $\lambda > \lout$, the dust is sufficiently warm everywhere that the
emission is truncated only by the matter distribution. The observed size is
then $\Theta$, the angular displacement of the source edge from the star, the
same for all wavelengths. However, when $\lambda < \lout$ the brightness is
truncated when $T(b) \le T_\nu$ before the edge of the source is reached,
resulting in a wavelength-dependent angular size $\theta_\lambda < \Theta$. The
observed size of optically thin emission {\em increases with wavelength\/} so
long as $\lambda < \lout$, the opposite of the trend at scattering wavelengths.

The frequency variation of the dust cross section is well described by
$\sigma_\nu \propto \nu^n$ with $n$ = 1--2. Then to a good degree of
approximation, the temperature variation of optically thin dust is $T \propto
1/r^t$, where $t = 2/(4 + n)$, producing the wavelength-dependent observed
angular size
\eq{\label{eq:size}
   \theta_\lambda = \Theta\x\cases{
   \left(\DS\lambda\over\lout\right)^{1/t}  & $\lambda < \lout$ \cr \cr
   1                                             & $\lambda \ge \lout$ }
}
Since $t < \half$, $\theta_\lambda$ increases faster than $\lambda^2$, a fairly
steep rise.

\subsection{Flux---the SED}

The flux can be obtained from equation \ref{eq:Basic} by integration over the
observed area. At emission wavelengths, the flux at distance $D$ is
\eq{\label{eq:flux}
    F_\nu = {\sigma_\nu\over D^2}\int B_\nu(T)\,\n\,dV.
}
Since the temperature profile of optically thin dust depends only on distance
from the star, the dependence on the source geometry enters only from \n.

As before, the integration is truncated by either the temperature or the matter
distribution. Whenever $\lambda > \lout$ at every point on the surface, the
emission is matter bounded everywhere and the integration encompasses the
entire source. Under this circumstances $F_\nu \propto \nu^2\sigma_\nu$, a
universal SED that depends only on the dust properties irrespective of
geometry. In particular, the spectral variation $\sigma_\nu \propto \nu^n$
gives $F_\nu \propto \nu^{2 + n}$, therefore the signature of matter bounded
emission is this SED accompanied by wavelength independent images; this is the
expected behavior in any geometry at sufficiently long wavelengths. At $\lambda
< \lout$ the integration volume is truncated by the temperature, and since the
emission volume decreases as the frequency increases, the rise of $F_\nu$ with
$\nu$ becomes less steep than in the matter dominated regime. Therefore, the
SED changes from $F_\nu \propto \nu^{2 + n}$ at $\lambda > \lout$ to a flatter
slope at $\lambda < \lout$.

The break in the slope at \lout\ can be used to determine the surface
temperature \Tout. Flux spectral variation shallower than $\nu^3$ is a clear
indication of temperature-bounded emission and should be accompanied by an
image size that increases with wavelength.

\subsection{Spherical Geometry}
\label{sec:Spherical}

Some explicit results are easily derived in the case of spherical symmetry.
Thanks to the scaling properties of dust radiative transfer (IE), only two
properties are required to specify the geometry. The first is the radial
optical depth at one wavelength, say $\tV = \sigma\sub{V}\int\n dr$ where
$\sigma\sub{V}$ is the cross-section at visual; at every other wavelength,
$\tau_\nu = q_\nu\tV$ where $q_\nu = \sigma_\nu/\sigma\sub{V}$. The second is
the dimensionless, normalized profile of the dust density distribution
\eq{\label{eq:eta}
    \eta(y) = {\n(y)\over\DS\int_1^{\infty}\!\!\!\!\n dy}
}
where $y = r/\Rs$; note that $\int\eta dy = 1$. Explicit results follow
immediately for all power-law density profiles where
\eq{\label{eq:power}
    \eta = {\N\over y^p}  \hskip 2cm
      \N = \cases{(p - 1)/(1 - Y^{1 - p})   & $p \ne 1$ \cr
                  (\ln Y)^{-1}              & $p = 1$}
}
The shell extends to the outer radius $Y\Rs$, subtending the angular region
$\ts \le \theta \le \Theta$, where $\ts = \Rs/D$ and $\Theta = Y\ts$. At
scattering wavelengths
\eq{\label{eq:SPHsca}
  I_{\nu,\rm sca}(\theta) =  \frac{\N}{2\pi }\tV\,L_\nu\varpi_\nu q_\nu
  \left(\ts\over\theta\right)^{p + 1}
}
\[\hskip 1.5in
 \x\int_0^{\sqrt{(\Theta/\theta)^2 - 1}}
  \!\!\!\! {du\over(1 + u^2)^{(p + 2)/2}}
\]
Whenever $\theta \ll \Theta$ the integration can be extended to $\infty$,
yielding $I(\theta) \propto 1/\theta^{p + 1}$; the brightness decreases as a
power law so long as the observation direction is not too close to the halo
edge. At emission wavelengths, on the other hand,
\eq{\label{eq:SPHemis}
  I_{\nu,\rm em}(\theta) =  {4\N\over c^2} k\Ts\,\tV\nu^2q_\nu
                   \left(\ts\over\theta\right)^{p + t - 1}
}
\[\hskip 1.5in
    \x\int_0^{\sqrt{(\theta_\lambda/\theta)^2 - 1}}
        \!\!\!\! {du\over(1 + u^2)^{(p + t)/2}}
\]
where the observed size $\theta_\lambda$ is smaller than $\Theta$ when $\lambda
< \lout$ (equation \ref{eq:size}). As long as $\theta \ll \theta_\lambda$ the
integration can be extended to $\infty$ and the brightness then decreases along
any radial direction in proportion to $1/\theta^{p + t - 1}$.

The flux integration in equation \ref{eq:flux} is similarly terminated at the
observed boundary $\theta_\lambda$, producing
\eq{
 F_\nu = {8\pi\N\over c^2}\,k\Ts\,\theta_{\rm s}^2\tV\
          {\nu^2 q_\nu\over 3 - (p + t)}\left[
          \left(\theta_\lambda\over\ts\right)^{3 - (p + t)} - 1\right].
}
Since $\theta_\lambda > \ts$, there are two families of SED. In the case of
steep density distributions with $p > 3 - t$, the first term inside the
brackets can be omitted because $3 - (p + t) < 0$. Such distributions produce
$F_\nu \propto \nu^2\sigma_\nu$ irrespective of the actual value of $p$. Since
typically $t \sim 0.4$, this behavior applies to all cases of $p \ga 2.6$. On
the other hand, whenever $p < 3 - t$ the omitted term dominates and the SED is
a broken power law. The power index switches from the universal $2 + n$ at
$\lambda \ge \lout$ to the geometry-dependent value ${3 + n - (3 - p)/t}$ (see
also Harvey et al 1991) at $\lambda \le \lout$.

These results show that the SED displays a dependence on the density profile
only when $p \la 2.6$, and then only in the spectral region $\lambda \le
\lout$. All other regions of $p$ and $\lambda$ produce the universal behavior
$F_\nu \propto \nu^2\sigma_\nu$.

\label{lastpage}
\end{document}